# Studies of DC biasing of internal ring on plasma rotation and transport in a toroidal geometry.

Å. Fredriksen[a], C. Riccardi[b], and S. Magni[b],

[a]Dep. Physics, University of Tromso, Norway
[b]Dipartimento di Fisica "G.Occhialini" and INFM
Universita' degli Studi di Milano – Bicocca, Italy

We report results from experiments performed to study how a DC bias with respect to the plasma potential is affecting the plasma states in the toroidal geometry of the Blaamann device in Tromso. In the experiments discussed here, a ring with smaller diameter than the limiter was centered inside the bulk of the plasma and its bias varied with respect to the plasma potential.

In the electron saturation current regime of the ring, a significant reduction of the fluctuation levels was observed, and a shear in the poloidal velocity occurred at the low-field side of the ring. This shear was positioned at the same radial position as the maximum of the radial transport when the ring was left floating or biased in the ion saturation current regime. For the latter ring biases, the poloidal velocities had no shear on the low-field side of the ring. Without the velocity shear, the radial transport was similar to previous experiments without ring. With the velocity shear, the radial transport was destroyed. This plasma state, with very low radial transport and fluctuation levels as well as plasma densities around $10^{17}$ m$^{-3}$, is believed to provide a suitable plasma for wave propagation studies in a magnetized plasma with curved magnetic field lines without end effects.

## Introduction

In the last decades, it has been confirmed that reduction of radial transport in fusion devices is closely linked to decorrelation of turbulent structures due to shears in electric fields and poloidal velocities [1-3]. Consequently, many studies have been performed in order to achieve better control of plasma confinement by edge biasing [4-6]. Also basic plasma experiments on plasma rotation of relevance to space plasmas and applications have been conducted recently in linear devices [7]. In this paper, we report effects from biasing an internal electrode in a plasma with a simple toroidal geometry, i.e. without a poloidal transform.

## The experiment

The simple magnetized torus Blaamann [8] has 24 magnetic field coils, which produce a toroidal magnetic field of up to 3.5 kG with ripple, $\delta B/B$, less than 0.01. The plasma is produced by a vertical hot tungsten cathode positioned 1-2 cm inside the center of the equatorial plane of the smaller cross-section of the torus. It is heated to electron emission and biased at -140 V relative to the grounded chamber walls, producing an energetic primary electron beam, which ionizes the neutral gas. The resulting plasma exhibits a negative potential well near the centre of the poloidal cross-section and is subject to vertical $\nabla B$ and curvature drifts and poloidal **E**x**B**-drift, and the degree of ionisation is typically

~ 1 %. A poloidal limiter extending two cm inwards from the walls receives the discharge current between the filament and the common ground. There is no toroidal current or poloidal magnetic field imposed on the plasma, and hence no poloidal transform exists.

The present experiment was carried out in a helium gas at a pressure of $1.0 \times 10^{-3}$ mbar and discharge current 1 A. The toroidal magnetic field was set to 572 G, and the filament was biased at –140 V with respect to the walls. An 8 cm diameter ring electrode made of 6 mm diameter stainless steel tube was inserted vertically in the poloidal cross-section next to the limiter, as shown schematically in Fig 1 a).

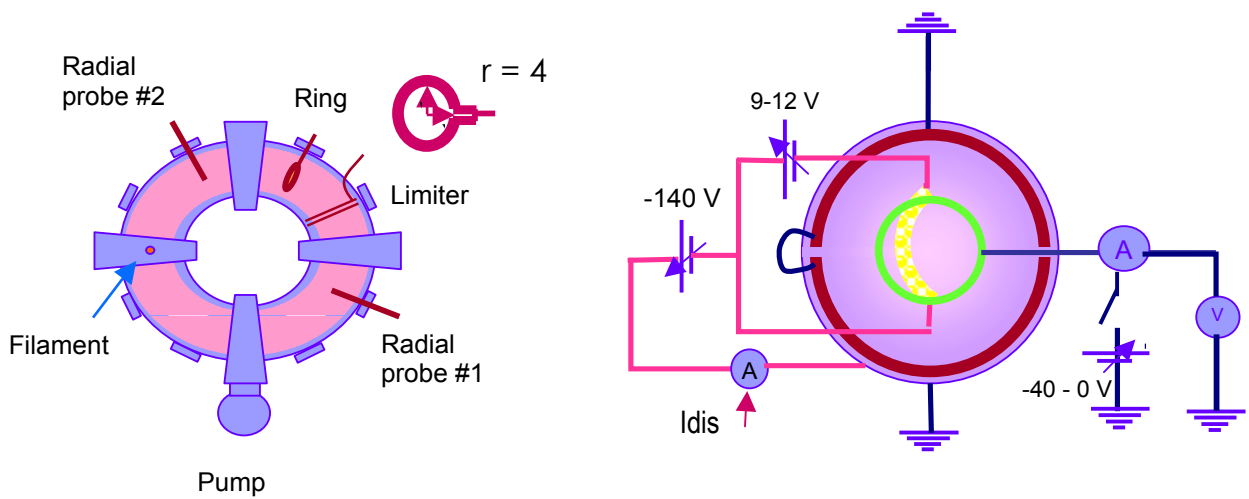

*Figure 1. Experimental layout with biased ring.*

The electrical circuit of the experiment including the filament circuit is shown in Fig. 1b). The discharge current between the filament circuit and the grounded chamber walls is about 1.0 A without the ring being connected to the circuit (i.e. floating ring). In this case, the ring is floating at -22 V. When the ring is grounded, it draws an electron current of about 650 mA, while the discharge current is reduced to 700 mA. In the following, we will discuss results from experiments when the ring is grounded (at zero V), floating, and biased at – 40 V, respectively.

The diagnostics was carried out by means of Langmuir probes, one probe obtaining swept Langmuir IV characteristics (probe #1) and a second 3-pin probe measuring radial flux parameters (probe #2). The vertical displacement was 2.5 mm for the probe pins measuring the poloidal E-field fluctuations, and the electron saturation current was sampled for density fluctuations at a probe pin placed vertically in the midpoint between the E-field pins.

Data was acquired by radially scanning the probes inwards from a position close to the limiter on the low-field side (outside the filament) to a position well inside the filament on the high-field side.

The plasma parameters derived from the Langmuir scans are shown in Fig. 2. Results with floating and grounded ring are plotted, with the limiter being grounded, as indicated in Fig. 1 b).

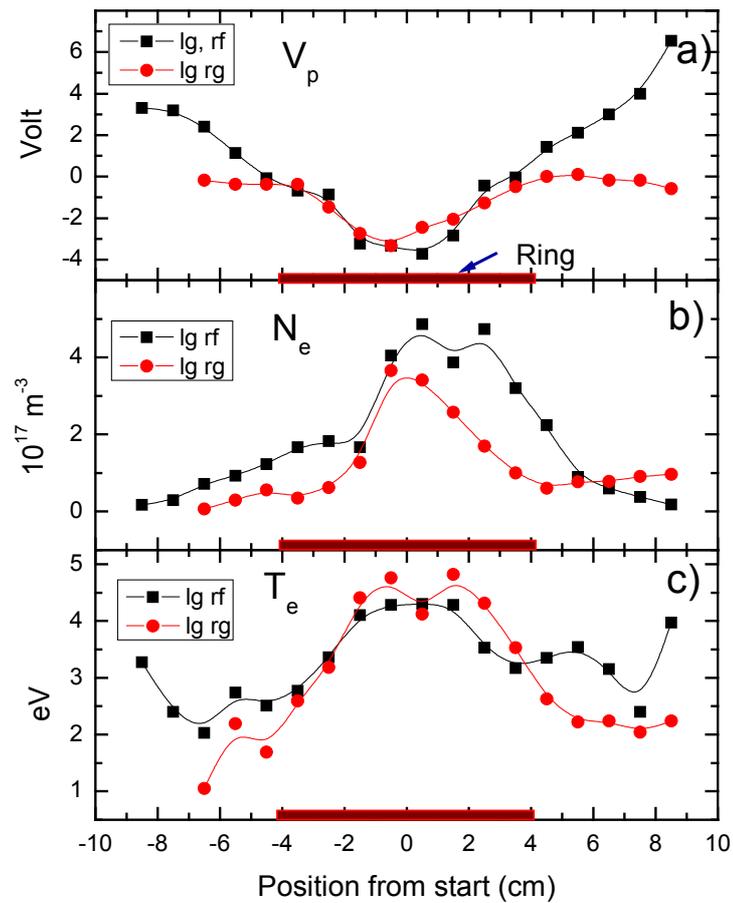

Figure 2. *Plasma potential a), electron density b) and electron temperature c) as a function of radial position, with the ring floating (black squares) and grounded (red circles). The thick red lines indicates the diameter and radial extend of the ring.*

With the ring floating, the profile of the plasma potential ($V_p$, Fig. 2a)) is similar to the usual case without any ring inserted in the cross-section, with the exception of a minor flattening of the profile near the circumference of the ring. On the other hand, outside the circumference of the grounded ring the plasma potential is nearly constant with values close to zero.

The electron density (Fig. 2 b) is skewed towards the low-field side of the circumference of the floating ring. In this region, the density is also higher than in the case of the grounded ring. This may be explained by the fact that the grounded ring is drawing an electron saturation current of about 650 A, and thus acts as an effective sink for the plasma particles. Also around the high-field side of the grounded ring, the density is depleted by nearly a factor two.

The electron temperature (Fig. 2c) is only slightly affected by the grounded ring, being less by approximately one eV outside the circumference of the ring in this case.

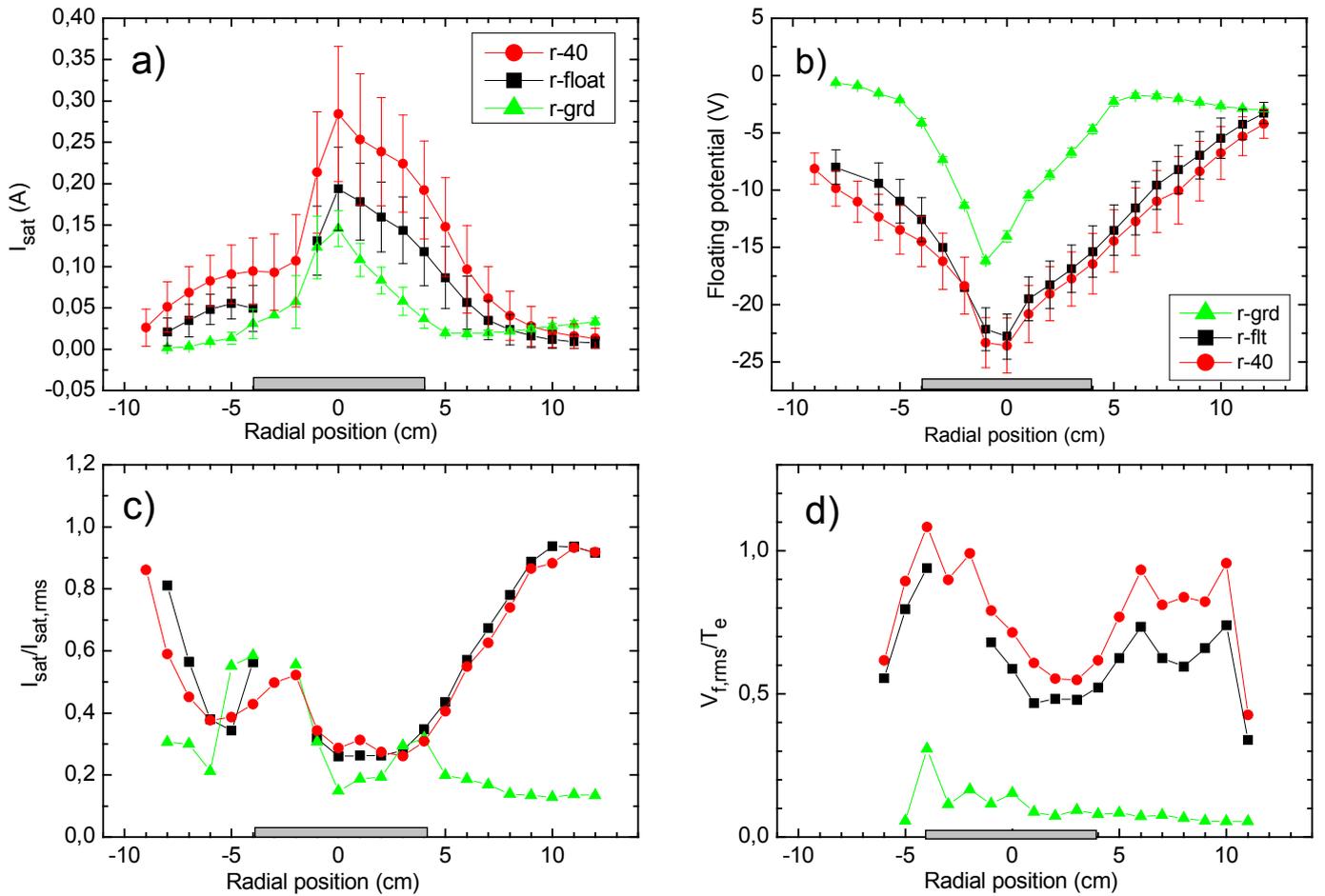

*Figure 3. Average values with standard deviation of electron saturation current (a) and floating potential (b), and of relative fluctuation levels of electron saturation current (c) and floating potential (d). Data shown is with ring bias -40V (red circles), floating (black squares), and grounded (green triangles). The grey line on the axes shows the radial extension of the ring.*

The average values of the electron saturation current ($I_{sat}$) shown in Fig. 3a) is increasing when the ring is biased at negative potential with respect to the floating potential, i.e. in the ion saturation regime. On the other hand, the floating potential ($V_f$) is not much affected by this ring bias (Fig. 3b), apart from a small increase in the relative fluctuations compared to the case with a floating ring (Fig. 3d). The relative fluctuations of $I_{sat}$ are not affected at all. This shows that a negative bias with respect to the plasma is improving the plasma confinement, but at a cost of slightly higher fluctuation levels in the floating potential, possibly because of an increased growth rate in the flute instability due to larger density gradients.

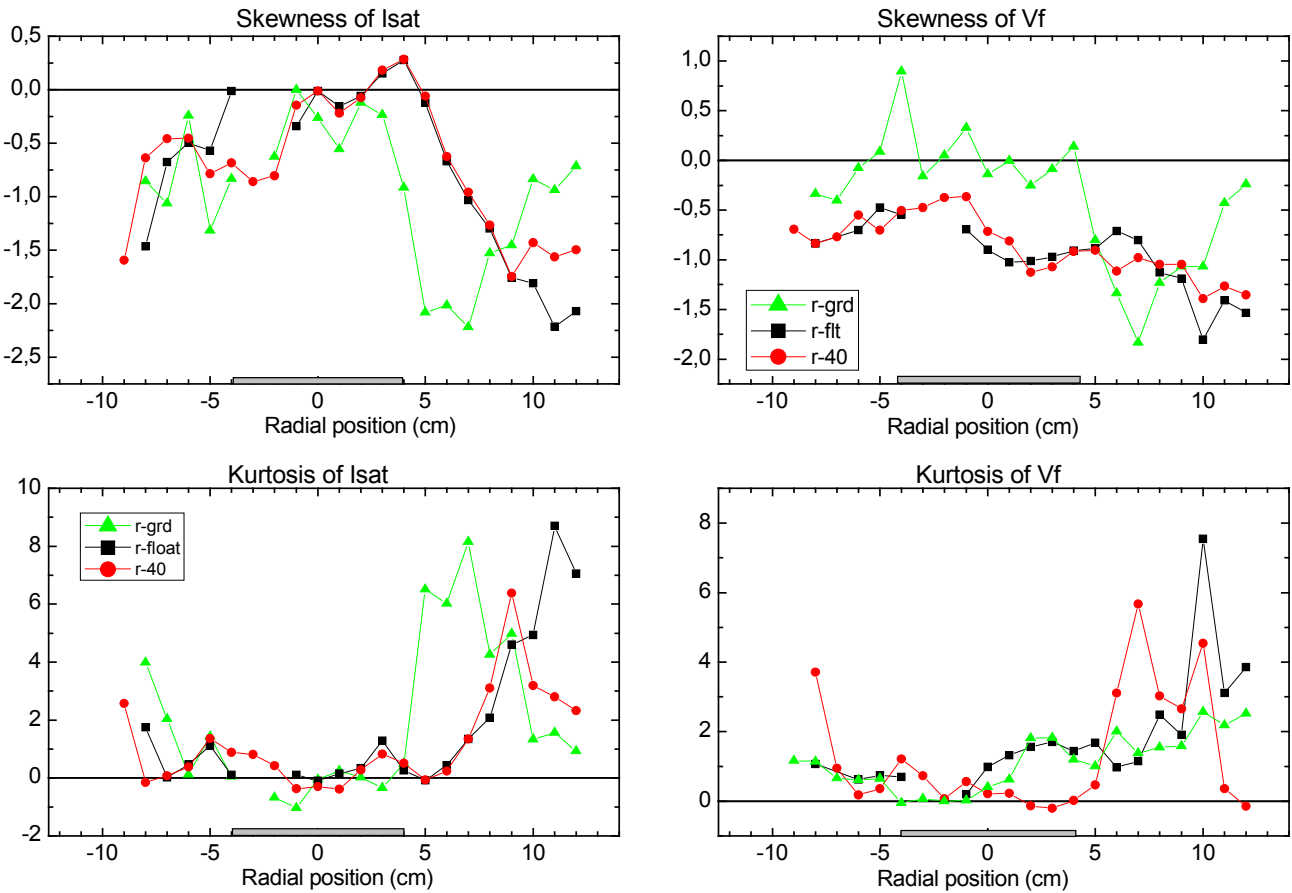

*Figure 4. Skewness of Isat( a) and Vf (b) and kurtosis of Isat (c) and Vf (d). Red circles correspond to ring biased at -40 V, black squares to ring floating, and green triangles represent data with grounded ring.*

On the other hand, when the ring is grounded, at a potential in the electron saturation regime, $I_{sat}$ is reduced below the level of the current when the ring is floating, as noted already from the Langmuir density measurements. However, as seen from the error bars in Fig 3a) and b), the fluctuations are significantly reduced in both $I_{sat}$ and $V_f$. The relative fluctuations are severely reduced, especially in the floating potential (Fig 3d), which is reduced to 0.1 or less on the low-field side of the filament. The relative fluctuation levels of $I_{sat}$ are reduced to less than 0.2 on the low-field side of the ring, but inside they are of the same order as the saturation current fluctuations in the floating ring case. The grounded ring thus produces a very quiescent plasma state in this device, which otherwise confines a very turbulent plasma. Although the plasma density near the center of the cross section is reduced somewhat, the densities are still of the order of $10^{17}$ m$^{-3}$ over radial span of more than 6 cm. The plasma potential profile has on the same time a very small gradient in this region. Assuming that this plasma state can be achieved for a set of different magnetic fields and neutral gas pressure, this configuration should provide excellent conditions for the systematic study of wave propagation in magnetized plasma with a curved geometry and without end effects.

The other statistics moments, i.e. skewness and kurtosis, are shown in Fig. 4. The skewness in $I_{sat}$ is now Gaussian inside the grounded ring, as is also the case for the other ring biases. On the low field side of the ring the skewness of Isat is sharply decreasing to large negative values in the grounded case, while it for the other cases reach the same skewness values more gradually towards the limiter.

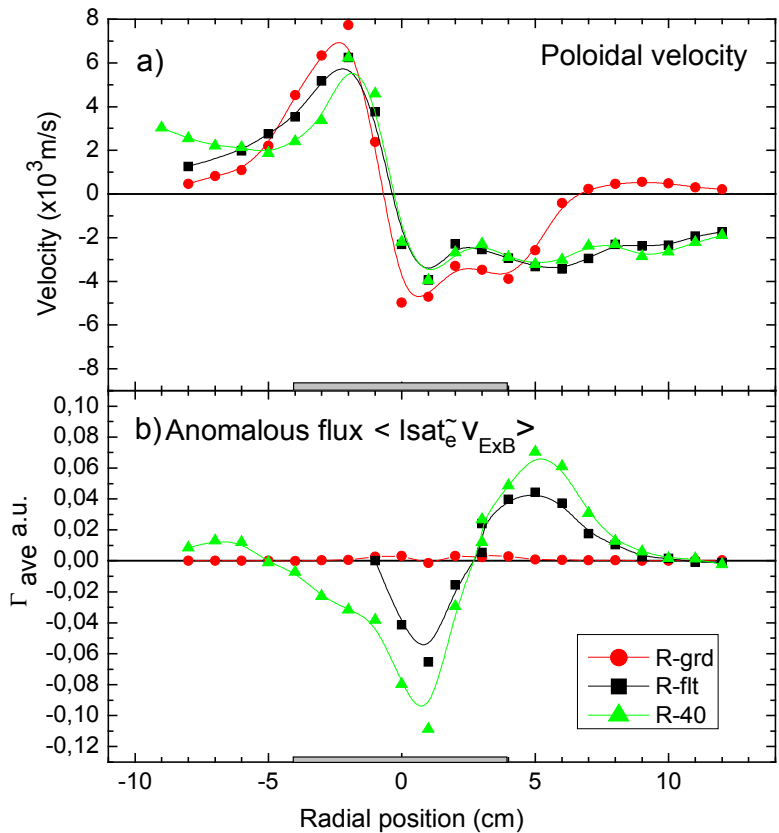

*Figure 5. Poloidal velocity (a) and turbulent radial flux (b). Red circles represent case with grounded ring, black squares case with floating ring, and green triangles are data with ring biased at -40 V.*

Overall, the floating potential becomes more Gaussian with The skewness of Vf is close to zero inside the grounded ring, while for the biased rings the skewness is negative throughout the radial cross section. For both the $I_{sat}$ and $V_f$ the kurtosis is close to zero inside the ring at all bias values, while it deviates more from zero outside the ring.

The poloidal velocities are plotted in Fig. 5a). It can be noted that for the ring biased at – 40 V or at floating potential, the plasma is rotating in the same sense as has been observed in previous experiments [9], with nearly constant negative velocity in the low-field side throughout the radial cross-section. On the other hand, when the ring is grounded, there is a strong velocity shear just outside the ring, bringing the rotation down to nearly zero velocity and slightly positive. It should be noted that this shear coincides with the location of maximum radial transport in the cases with floating and negatively biased ring. For these biases the plasma rotation is similar to what is commonly observed in the Blaamannn plasma, i.e., constant throughout the cross section of the low-field side. Thus, there is strong evidence that the velocity shear imposed at the position of maximum radial transport is destroying the radial transport.

**Conclusion.**

We have performed experiments to study how a ring biased with respect to the plasma potential is affecting the plasma states in the toroidal geometry of the Blaamann device in Tromso. A ring with smaller diameter (8 cm) than the limiter was centered inside the bulk of the plasma and its bias varied with respect to the plasma potential.

In the electron saturation current regime of the ring (at ground potential), a significant reduction of the fluctuation levels was observed, and a shear in the poloidal velocity occurred at the low-field side of the ring. This shear was positioned at the same radial position as the maximum of the radial transport when the ring was left floating or biased in the ion saturation current regime. For the latter ring biases, the poloidal velocities had no shear on the low-field side of the ring. Without the velocity shear, the radial transport was similar to previous experiments without ring. With the velocity shear, the radial transport was destroyed. Very low radial transport and fluctuation levels as well as plasma densities being still around 1017 m-3, characterize this plasma state. It is believed to provide suitable plasma for wave propagation studies in magnetized plasma with curved magnetic field lines and no end effects.

**References.**
[1] Burrell. K. H., Phys. Plasmas, 4 (1997) 1499
[2] Conway et al., Phys. Rev. Lett., 84 (2000) 1463
[3] Hidalgo, C. et al., Nuclear Fusion, 31 (1991) 1471
[4] Magni S., Riccardi, C., and Roman, E., Phys. Plasmas, 11 (2004) 4564
[5] Mase, A., et al., Nuclear Fusion, 31 (1991) 1725
[6] Cabral, J. A. C., et al., Plasma Phys. Contr. Fusion, 40 (1998) 1001
[7] Shinohara, S., Matsuoka, N., Yoshinaka, T., Jap. J. Appl. Phys. 38 (1999) 4321
[8] Rypdal, K. et al., Plasma Phys. Contr. Fusion, 36 (1994) 1099
[9] Riccardi, C., and Fredriksen, Å., Phys. Plasmas, 8 (2001) 199